\documentclass[fleqn,twoside]{article}
\usepackage[headings]{espcrc2}

\readRCS
$Id: espcrc2.tex,v 1.2 2004/02/24 11:22:11 spepping Exp $
\usepackage{graphicx}
\usepackage[figuresright]{rotating}

\newcommand{\AmS}{{\protect\the\textfont2
  A\kern-.1667em\lower.5ex\hbox{M}\kern-.125emS}}

\title{Recent Results from RHIC \& Some Lessons for 
Cosmic-Ray Physicists}

\author{Spencer R. Klein\address[LBNL]{Nuclear Science Division,
Lawrence Berkeley National Laboratory, Berkeley, CA, 94720, USA}
        \thanks{Email: srklein@lbl.gov}
	}

\runtitle{RHIC Results}
\runauthor{S. Klein}

\begin{document}

\begin{abstract}

The Relativistic Heavy Ion Collider (RHIC) studies nuclear
matter under a variety of conditions.  Cold nuclear matter
is probed with deuteron-gold collisions, while hot nuclear matter
(possibly a quark-gluon plasma (QGP)) is created in heavy-ion 
collisions.  The distribution of spin in polarized nucleons is measured with
polarized proton collisions, and photoproduction is studied using
the photons that accompany heavy nuclei.  

The deuteron-gold data shows less forward particle production than
would be expected from a superposition of $pp$ collisions, as expected due
to saturation/shadowing.  
Particle production in $AA$ collisions is well described by a model of
an expanding fireball in thermal equilibrium.  Strong hydrodynamic
flow and jet quenching shows that the the produced matter
interacts very strongly.  These phenomena are consistent with new
non-perturbative interactions near the transition temperature to the QGP.

This writeup will discuss these results, and their implications for
cosmic-ray physicists.
\end{abstract}

\maketitle

Since the first collisions in 
2001, RHIC has produced a wealth of data on $pp$, $d$Au,
AuAu and CuCu collisions at center of mass energies from 20 to 
200 GeV per nucleon pair. RHIC was built to study aspects of QCD;
the main foci have been the study of 
cold nuclear matter via dA collision, and of hot nuclear matter (the Quark 
Gluon Plasma (QGP)?),
via heavy-ion collisions.  RHIC also studies polarized proton collisions
to measure the polarized parton distributions, and photonuclear
interactions.  This data is also of interest for
modelling cosmic-ray air showers.   

RHIC hosts 5 experiments.  There are two large experiments,
PHENIX and STAR, and 3 smaller ones:  PHOBOS, BRAHMS and pp2pp.

PHENIX has two large central spectrometers,  and a forward muon 
system \cite{PHENIX}.  It is optimized for particle identification, 
particularly for leptons.

STAR is optimized for global event studies, with a large acceptance
time projection chamber for charged particles and a calorimeter to
detect neutral particles \cite{STAR}.  

PHOBOS detects charged particles over a very large pseudorapidity range, $|\eta|<5.4$,
and has two small spectrometer arms for tracking \cite{PHOBOS}.  Pseudorapidity
$\eta= -\ln{[\tan(\theta/2)]}$, where $\theta$ is the angle between the particle
direction and the beam axis.  BRAHMS
consists of central and forward spectrometers, with precise particle
tracking and identificaton in a small solid angle \cite{BRAHMS}.  pp2pp
consists of Roman pots which track protons scattered at small angles; it 
studies $pp$ diffraction \cite{pp2pp}.

This writeup will review the different RHIC physics, starting with polarized 
proton collisions and photoproduction, before moving on to discuss cold and 
hot nuclear matter.

\section{$pp$ and Polarized Proton Collisions}

$pp$ collisions serve two functions at RHIC:  tests of pQCD
calculations with unpolarized 
$pp$ collisions, and measurements of the polarized parton 
distributions.

\begin{figure}[htb]
\includegraphics[width=2.75in,clip]{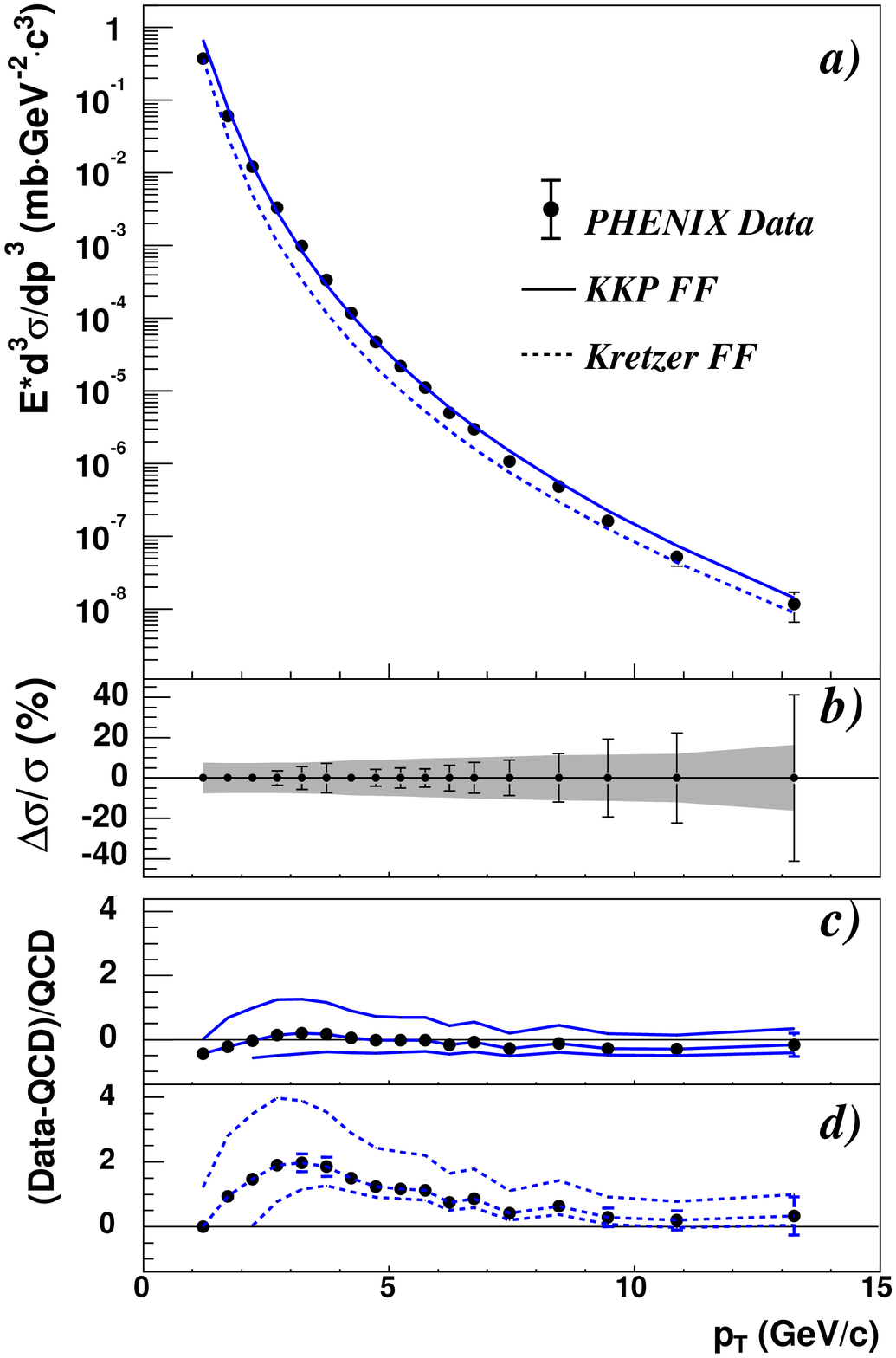}
\includegraphics[width=2.75in,clip]{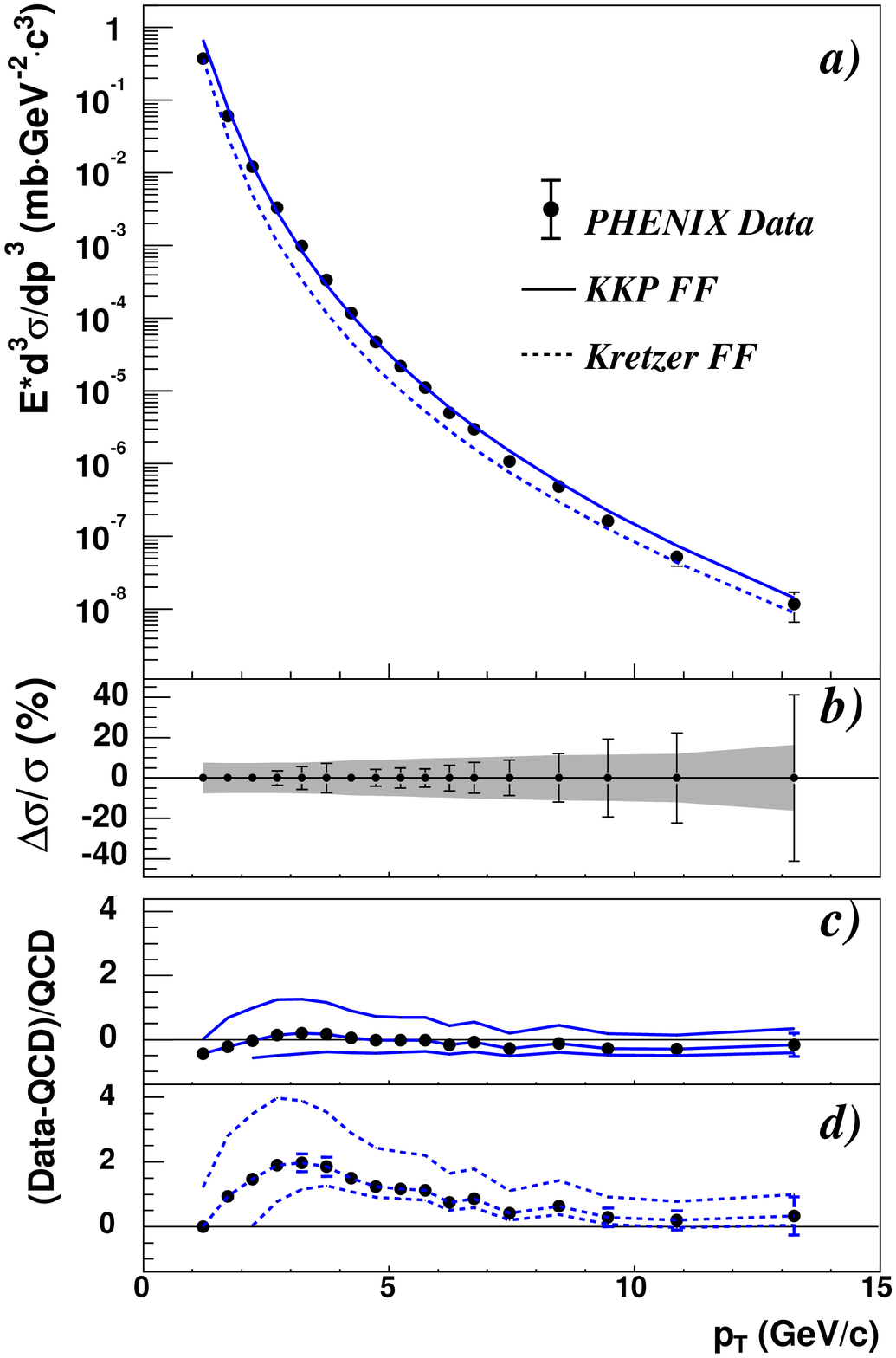}
\caption{PHENIX data on $\pi^0$ production in 
in 200 GeV $pp$ collisions, compared with NLO pQCD calculations plus
two different fragmentation functions \cite{PHENIXpi0}.} 
\label{fig:ppQCD}
\end{figure}

Jet and single particle cross section data at RHIC are in good agreement
with recent perturbative QCD calculations \cite{PHENIXpi0,STARspin}.
Figure \ref{fig:ppQCD} compares pQCD calculations
with data for $\pi^0$ cross sections; 
the agreement is good.  This good agreement is possible
thanks to recent next-to leading order pQCD calculations and 
improved fragmentation functions.

Polarized proton collisions are used to study the spin structure of the
nucleon \cite{spinreview}.  The total nucleon spin is the sum of the quark spins, 
gluon spins and orbital angular momentum within the proton.  
Experiments at SLAC and CERN in the 1970s-1990s showed that the
quarks carry a relatively small fraction of the nucleon spin; this is
sometimes called the 'spin crisis.'  RHIC will
measure the gluon polarization, via quark-gluon and gluon-gluon interactions.
The polarized gluon distributions are found by comparing the normalized
rates for collisions where the  proton spins are pointing in the same vs. 
in opposite directions, 
Although much more data is needed, already it is possible to exclude
the most extreme models of gluon polarization \cite{STARspin}. 

\section{Photoproduction}

Photoproduction has been studied in gold-gold and deuteron-gold collisions.
Virtual photons from the electromagnetic field of one nucleus interact
with the other nucleus.
Photoproduction of the $J/\psi$ is sensitive to the gluon distributions
of the target nucleus \cite{nystrand}.  $\rho$ and $\pi^+\pi^-\pi^+\pi^-$ 
final states have also been studied \cite{kleindis05}.

Vector meson photoproduction can occur
two ways: nucleus 1 can emit a photon which interacts with nucleus 2,
or vice-versa.  The two channels are indistinguishable, so they interfere.
Going from one emitter to the other is a parity inversion, and
vector mesons are negative parity, so the two amplitude subtract.
At mid-rapidity
\begin{equation}
\sigma(b) = \sigma_0(b) (1-\cos(p_T\cdot b)).
\end{equation}
Here, $\sigma_o$ is the cross section without interference
the $p_T$ is that of the vector meson, and $b$ is the impact parameter
(distance between the two ion centers at closest approach).
$b$ is not an observable, so the total cross section is the integral
of Eq. 1.  This interference has been observed through a 
reduction in cross section for $p_T < \hbar/\langle b\rangle$ \cite{qm2004}.  

Because of the strong nuclear fields, some reactions can uniquely be studied
in heavy-ion collisons.  At RHIC, multi-photon
interactions involving a single ion pair {\it i.e.} $ Au + Au \rightarrow
Au^* + Au^* + \rho$ via 3 photon exchange (one for each
nuclear exchange, and a 3rd to produce the $\rho^0$) has also been studied \cite{kleindis05}.
The LHC will reach $\gamma p$ center of mass energies up to several hundred GeV, 
well beyond the reach of HERA \cite{upcreview}. 

\section{Cold Nuclear Matter}

By comparing
$pp$ and $dA$ collisions the effects of the nuclear environment
may be studied.  The nuclear environment is expected to alter parton densities,
possibly leading to qualitatively new behavior \cite{mark}.  Gluons have a virtuality ($Q^2)$ dependent 
transverse area $\pi(\hbar c)^2/Q^2$. At high densities (i.e. at low $x$),
partons will recombine,  in reactions like $ g + g \rightarrow g$.  This 
recombination moderates the growth
in gluon density as $x$ decreases.  Because of the higher parton density, recombination is more
significant in nuclei, and comparisons of parton distributions in protons and in
nuclei are sensitive to these changes; the difference measured in lepton-nucleon 
interactions is known as shadowing.

\begin{figure}[tb]
\includegraphics[width=3in,clip]{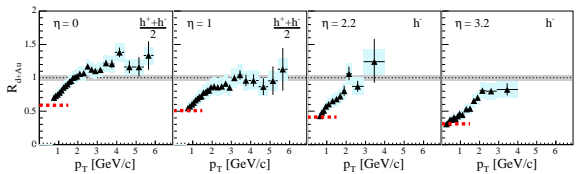}
\caption{The ratio of charged particle production in $dA$ and
$pp$ collisions (normalized to the number of nucleon-nucleon collisions)
at different pseudorapidity.  In the absence of nuclear effects, $R_{dAu}$ should
equal 1. \cite{BRAHMSrda}.}
\label{fig:BRAHMS}
\end{figure}

Figure \ref{fig:BRAHMS} shows the effect of this density on particle production.  It 
compares charged-particle production in $dAu$
and $pp$ collisions at different pseudorapidity \cite{BRAHMSrda}.  Higher pseudorapidity corresponds to
higher-$x$ partons from the deuteron, and lower-$x$ partons from the gold; as expected
from recombination, shadowing is larger at lower-$x$.

Different theoretical studies have
used BFKL and/or DGLAP evolution, or taken advantage
of the relationship between shadowing and diffraction.  One interesting
new approach, the colored glass condensate (CGC) treats the
gluon fields in the nucleus as a classical field.  

\begin{figure}[tb]
\includegraphics[width=2.75 in,clip]{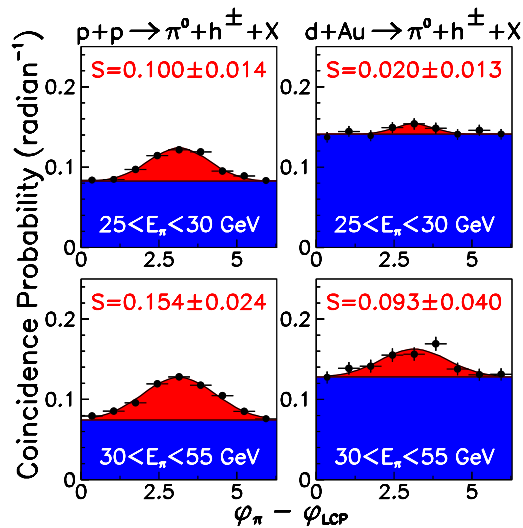}
\caption{Azimuthal correlations between a forward $\pi^0$ and a more
central charged hadron in (left) $pp$ and (right) $dA$ collisions for
(top) $25 < E_{\pi} < 30$ GeV and (bottom) $30 < E_{\pi} < 55$ GeV \cite{STARdA}.  
The lower (blue) area shows the background, while the higher (red) area follows
a fit to the data (points).  The 'S' values give the size of the peak, smaller
in $dA$ collisions.  The correlations 
are expected in dijet events.}
\label{fig:2part}
\end{figure}

The CGC makes some interesting predictions \cite{raju}; a CGC should
interact coherently as a single object, producing some new effects.
When a parton interacts with a CGC, the CGC recoils coherently; because of it's high mass, 
the recoil is muted, leading to apparent 'monojets'.  Without a CGC, 2 parton $\rightarrow$
2 parton reactions like $g + g \rightarrow g + g$ produce two azimuthally back-to-back
jets.  However, if the gluon strikes a CGC, the heavy object will recoil slowly, and
the recoiling gluon will produce the single visible jet. This process is studied
experimentally via 2-particle correlations.  Figure \ref{fig:2part} shows 
the azimuthal angle correlations between a $\pi^0$ produced in the forward direction
and a charged hadron produced near mid-rapidity, for $pp$ and $dA$ collisions.  Dijet
events should produce back-to-back correlations.  These
correlations are smaller for $dA$ collisions than for $pp$ collisions, as expected from
CGC models.  The suppression rises at smaller $\pi^0$ energies, as expected from a CGC.

\section{Hot Nuclear Matter}

A main experimental
goal of RHIC is to search for the quark gluon plasma (QGP), an interacting system of partons
in equilibrium \cite{QGPreview}.  Lattice gauge theory predicts that a QGP is produced when nuclear matter
at low baryochemical potential (baryon density) is heated to a temperature above about 170 MeV \cite{lattice}.

RHIC heats nuclear matter by colliding heavy ions.  The collisions occur in several
stages.  Initially, the ions collide and their partons interact, producing new partons.
These partons themselves interact, producing still more partons.  As these interactions occur, the system expands
and cools.  Eventually, the partons form hadrons which may themselves interact.  Interactions
continue until the system cools enough that no further inelastic processes are possible; this transition
is known as chemical freezeout.  Later, the hadrons separate enough that no further elastic collisions
occur; this is thermal freezeout.

Much RHIC data is analyzed in terms of centrality, or impact parameter ($b$).  
Several methods are used to characterize centrality:  $N_{part}$ is the number of nucleons participating
in the collision; for a head-on collision, $N_{part}=2A$.  $N_{bin}$ is the number of binary
(nucleon-nucleon) collisions; this is relevant for comparing pQCD participle production between
$pp$ and $AA$ collisions.   For a given $b$, $N_{part}$ and $N_{bin}$ are determined using
a Glauber calculation.  Centrality is also  given in percentages, such
as the $0-10\%$ most central collisions. 

\subsection{Overall Event Structure}

\begin{figure}[tb]
\includegraphics[width=2.65in,clip]{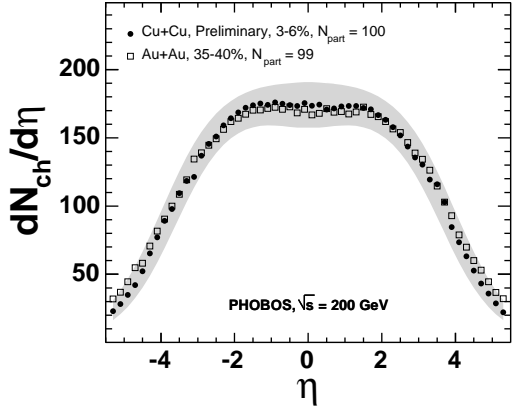}
\caption{$dN/d\eta$ for $3-6\%$ copper-copper collisions and $35-40\%$ gold-gold
collisions.  The two datasets have very similar $N_{part}$, and almost identical
$dN/d\eta$ \cite{gunther}.}
\label{fig:dndeta}
\end{figure}

We begin by considering the particle production.  Figure
\ref{fig:dndeta} compares the pseudorapidity ($\eta$) distribution for $3-6\%$ central copper-copper 
collisions with $35-40\%$ gold-gold collisions; the two datasets have almost identical
$N_{part}$.   At least for identical nuclei, $N_{part}$ determines the collision dynamics.
It can be used to model other systems, such as collisions involving nitrogen.

\begin{figure}[tb]
\includegraphics[width=3in,clip]{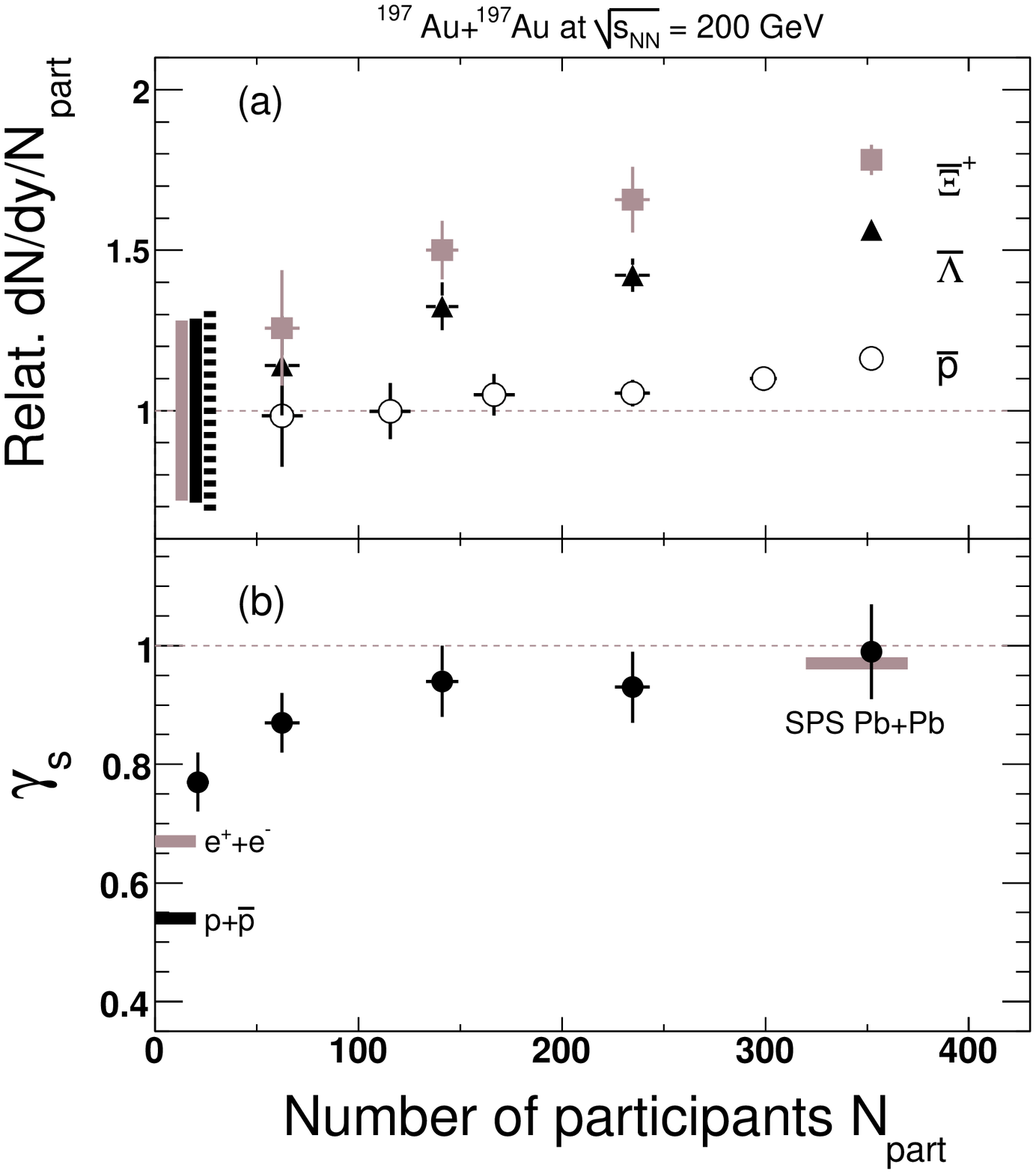}
\caption{Strangeness suppression factor $\gamma_s$ as a function of 
$N_{part}$. $\pi$, $K$, $p$, $\Lambda$, $\xi$ and $\Omega$ particles
and their antiparticles were used to determine $\gamma_s$ \cite{STARstrangeness}.}
\label{fig:strange}
\end{figure}

The final state composition is well described
by a thermal model, with production at an equilibrium temperature of
$T_c = 165\pm 10$ MeV; The abundance of particle species x, $N_x$
depends on its mass
$m_x$ \cite{STARstrangeness}:
\begin{equation}
N_x\approx exp{(-m_x/kT_c)}.
\label{eq:thermal}
\end{equation}
In contrast to $pp$ collisions, strangeness is
fully equilibrated; there is no strangeness suppression.  The system is described 
by a grand-canonical ensemble; a large thermal bath conserves strangeness, and individual
strange particles can be produced, rather than the pairs required in smaller systems.
Figure \ref{fig:strange} shows how the strangess suppression factor $\gamma_s$ varies
with $N_{part}$.  Here, $\gamma_s=1$ corresponds to thermal equilibrium, Eq. (\ref{eq:thermal}).
For $pp$ collisions, $\gamma_s\approx 0.54$.  Strangeness production is enhanced over 
$pp$ collisions even for relatively small systems, such as nitrogen-nitrogen collisions.
For heavy systems, the strangness content almost doubles.

A more detailed model of the collision includes both thermal energy of 
collective expansion, {\it i.e.} a hydrodynamically expanding fireball.  
The `blast wave' model describes this
hydrodynamic expansion
with two parameters: system temperature, $T$ and a collective expansion
velocity, $\langle\beta\rangle$, the velocity at the
outer edge of the fireball \cite{blast}.  Different velocity profiles can be used, with small
effects on the final results. Because of the collective expansion,
heavier particles have higher $\langle p_T\rangle$.
Fits to the $p_T$ spectra of different particles find $T\approx 106$ MeV, and
$\langle\beta\rangle\approx 0.55$c.  This model has been fit to a large body
of RHIC data, with considerable success \cite{blast}. It described soft-particle
production in ion collisions, and might be of use for simple air-shower simulations.

Elliptical flow is another aspect of the hydrodynamical behavior of the system.
In a non-central heavy-ion collision, the overlap region is
roughly almond shaped.  Pressure converts this spatial anisotropy
into a particle density anisotropy. The particle flux depends on
$\phi$, the azimuthal (perpendicular to the beam) angle
with respect to the reaction plane (line between the
centers of the two nuclei).  The dependence is
\begin{equation}
{dN\over d\phi} = 1 + 2 v_2\cos(2\phi)+...
\end{equation}
The elliptical flow, $v_2$ varies
with particle species and $p_T$. Additional terms for directed flow
(e.g. $v_1$) and a small quadrupole moment are ignored here.   Hydrodynamic
flow is a powerful diagnostic tool, since it probes particle interactions
(e.g. pressure) very early in the collision.    At low momenta, $p_T < 2$ GeV, the
measured particle flow is consistent with hydrodyamic models.  
The
initial-state spatial anisotropy is completely converted into a particle asymmetry.
Nuclear matter acts like a nearly perfect fluid.

\begin{figure}[tb]
\includegraphics[width=3in,clip]{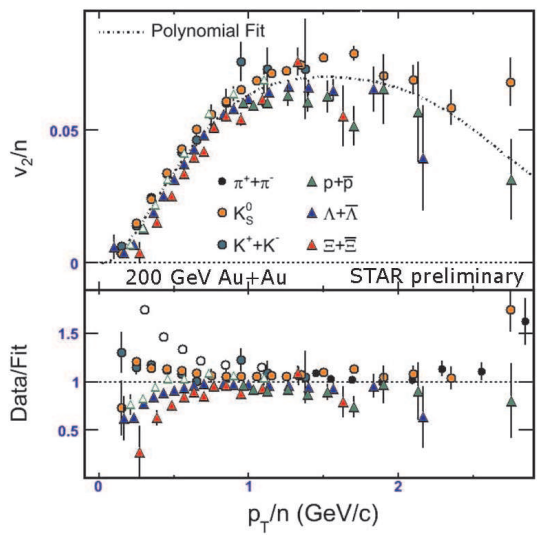}
\caption{Elliptic flow per constituent quark, $v_2/n$, ($n$ being the number of constituent quarks
in the hadron) vs. $p_T/n$ for various hadrons.  Except for the $\pi$, the flow of both mesons
and baryons follows a very similar line, indicating that quarks flow, rather than hadrons
\cite{lacey}.}
\label{fig:flow}
\end{figure}

Figure \ref{fig:flow} shows the elliptic flow per constituent quark, $v_2/n$
($n=2$ for mesons; $n=3$ for baryons), as a function of $p_T$ per constituent quark,
for a variety of different particles. Except for pions, all of the baryon and meson
data lie near a single line, with very similar per-quark flow \cite{lacey}.  It appears
that partons are flowing, rather than hadrons. 
Proposed explanations for the higher pion $v_2$ include their low mass, and/or feed-down 
from decays of heavier resonances.
Still, the $v_2$ data shows fairly clearly that the early interactions involve partons, 
rather than hadrons.

A final characteristic of the overall event is the system size at thermal freezeout.  This
is determined using Hanbury-Brown Twiss (HBT) interferometry, a measurement of the system
size that relies on the enhancement
in boson ($\pi$, etc.) production at small momentum differences, 
$|\vec{p}_1 - \vec{p}_2| <\hbar/R$.
By measuring the increased particle production at small momentum differences, the source size can be inferred. At
freezout (last interaction), central gold-gold collisions have a Gaussian radius of about 6 fm,
about twice the size of the original system.

\subsection{Perturbative Probes of the QGP}

\begin{figure}[tb]
\includegraphics[width=3in,clip]{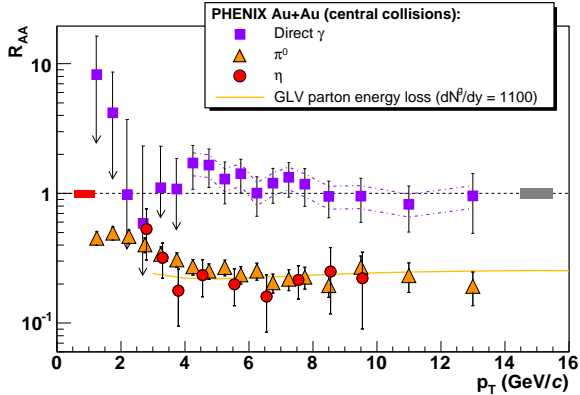}
\caption{$R_{AA}$ for $\pi^0$, $\eta$ and direct photons, together with a theoretical
calculation of energy loss for a plasma with a gluon density of 1100 gluons/unit of 
rapidity \cite{phenixraa}.  Hadrons with a color charge are heavily suppressed; photons
are not.}
\label{fig:phenixraa}
\end{figure}

Other studies of the QGP use high $p_T$ probes produced in the collisions.
High $p_T$ particle production is well described by pQCD.  Except for
some relatively minor nuclear effects, high $p_T$ (usually $p_T > 2$ GeV/c) 
particle production should be the same in
$pp$, $pA$ and $AA$ collisions, described by initial state parton distributions,
pQCD, and universal fragmentation functions.
Any large differences between systems
should be due to the interactions between the produced particles
and the nuclear medium. The time scale for fragmentation (whereby partons 
fragment into hadronic jets) is relatively long compared to the time a parton remains
in the fireball, so medium interactions should involve the produced
parton, rather than the final-state hadrons.  Energy loss in the medium will manifest
itself as a reduction in high $p_T$ particle production as energetic particles are
shifted to lower $p_T$.  This reduction is measured by comparing spectra from
central $AA$ collisions with appropriately scaled $pp$ and/or peripheral $AA$ collisions,
using the ratio
\begin{equation}
R_{AA} = {\sigma_{AA} \over N_{bin}\sigma_{pp}}.
\end{equation}
In the absence of nuclear effects, $R_{AA} = 1$.  Figure \ref{fig:phenixraa} shows $R_{AA}$ 
for $\pi^0$, $\eta$ and direct $\gamma$.  For direct photons $R_{AA}\approx 1$; 
photons do not interact with the medium.  However, for both types of hadrons, $R_{AA}\approx 0.2$ for 
$2 < p_T < 20$ GeV/c, showing a large nuclear suppression.
In $dAu$ collisions (not shown here),
$R_{AA}\approx > 1$, as expected due to initial state parton scattering.

In perturbative QCD, the suppression 
depends on the parton density in the fireball and the parton-parton cross section.  For the 
standard pQCD cross sections,
the observed energy loss requires a parton density of 
1100-1200 gluons/unit of rapidity \cite{vitev}.
This is far higher than the density inferred from the final state multiplicity.

$R_{AA}$ has also been measured for heavy quarks.  Since heavy quarks have a lower velocity
(for a given momentum) than lighter quarks, less radiative energy loss is expected.  However,
$R_{AA}$ for heavy quarks is similar to that for light quarks \cite{heavyquarks}.
This is difficult to understand in any pQCD calculation.  

$R_{AA}$ has also been measured for the $J/\psi$. For central AuAu collisions,
$R_{AA}\approx 0.3$ \cite{jpsi}. This is integrated over all $p_T$, so is not directly
comparable to the other $R_{AA}$ measurements. The suppression is comparable
to what was seen at $\sqrt{s_{NN}}=17.3$ GeV at the CERN SPS; this energy independence
is somewhat surprising.   

\begin{figure}[tb]
\includegraphics[width=3in,clip]{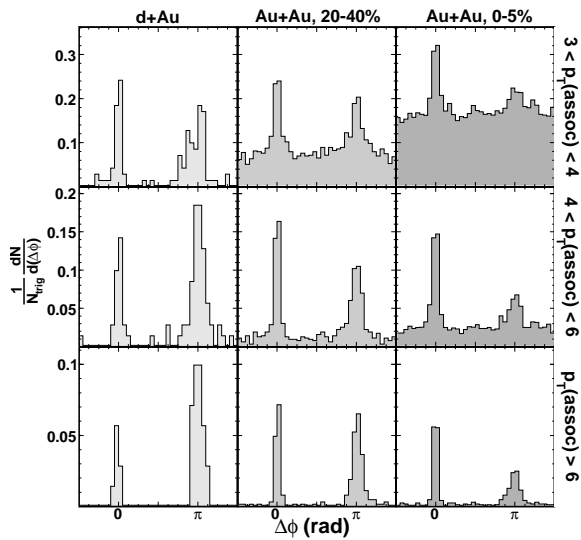}
\caption{Azimuthal correlations between a trigger particle with $p_T > 8$ GeV/c and
an associated particle in different $p_T$ ranges (in GeV/c) for $dAu$ and mid-peripheral and
central $AuAu$ collisions \cite{STAR2particle}.}
\label{fig:2particle}
\end{figure}

Parton energy loss can also be studied with
particle correlations. A high $p_T$ particle is selected as a 'trigger'
particle, and the azimuthal correlation with a lower $p_T$ 'associated' particle studied.
Figure  \ref{fig:2particle} compares the azimuthal correlations for $dA$ and for mid-peripheral
and central $AuAu$ collisions, for trigger particles with $p_T > 8$ GeV/c  and different
$p_T$ associated particles.  The $dAu$ data has a narrow peak for $\Delta\phi\approx 0$ (near-side) 
and a 
slightly broader peak for $\Delta\phi\approx\pi$ (far-side).  The former is from same-jet correlations, 
while the latter is from correlations involving two back-to-back jets;
the $pp$ data (not shown) has a similar structure. The mid-peripheral $AuAu$ data has a similar
behavior, although with higher backgrounds. In central $AuAu$ collisions, the back-to-back
peak is much smaller, especialy for softer associated particles.  The near-side
($\Delta\phi\approx 0$) peak is largely unchanged.  For softer trigger particles, the
far-side peak largely disappears.  

This data fits a model where the high $p_T$ particles
come from parton interactions near the surface of the fireball. 
Partons produced deeper in the fireball lose most of their energy by $dE/dx$, and so do
not produce high $p_T$ hadrons.   High $p_T$ partons produced near
the surface, pointing outward produce full-scale jets, with 'normal'
near-side correlations.  The suppression factor $R_{AA}$ depends on the surface to volume
ratio of the system. Back-to-back jets occur only for a narrow range of geometries, where 
a near-surface interaction produces back-to-back partons nearly tangential
to the surface.

\section{Recent Theoretical Developments}

The blast wave model does an excellent job of
modelling soft particle production at RHIC.  However, the fit parameters
imply very high interaction cross sections; these cross sections do not agree
with expectations from pQCD based particle interactions.  Several related puzzles
are also of interest:

1) The elliptic flow is at the hydrodynamic limit, with the produced matter behaving
like an almost perfect fluid.

2) The $R_{AA}$ measured at high $p_T$ can only be explained in a pQCD framework by an
unphysically high local parton density.

3) The $R_{AA}$ for heavy quarks is similar to that for lighter particles.

These puzzles have led to considerable theoretical speculation.  
Ed Shuryak has proposed that the temperature range $T_C < T < 4 T_c$
is a strongly coupled regime for partons \cite{Shuryak}; In this non-perturbative phase, there
are many very weakly bound colored states/resonances, such as $qq$, $gg$, $qqg$, etc.
These states are very lightly bound, so have large radii, leading to rescattering
cross sections 10 to 100 times those predicted by pQCD.  
Similar behavior is seen for atoms that have been tuned (via a magnetic field)
to be barely bound.  Shuryak also  pointed out that this strongly coupled QGP (sQGP)
may be expected based on duality arguments with weakly coupled string theory.  At
the same time, lattice calculations indicate that meson bound states (notably
including the $J/\psi$ surrvive up to temperatures considerably above
$\approx 1.5T_c$; this could explain the lack of additional suppression.

These strong interactions may explain the large elliptic flow
and reduced $R_{AA}$.  The strong interactions could also reduce the QGP lifetime
(as is seen in some HBT studies).  In short, this strongly-coupled phase
could explain many of these puzzling observables.  Of course, detailed quantitative
studies are needed.

One measure of the cross section is the fluid viscosity; elliptic flow depends
on the shear viscoity/entropy ($\eta/s$).  Flow data shows
$\eta/s < 0.1$ \cite{viscosity}, indicating that the 
sQGP is a 100 times better fluid than water.
This viscosity is much lower than
that expected for a hadron gas or
from a perturbative (i.e. hotter) QGP, and appears beyond the reach of perturbative
QCD.   In fact, it approaches the quantum limit, $\eta/s\approx 1/4\pi$, calculated 
using duality arguments \cite{viscosity}.

\section{Lessons for Cosmic-Ray Modelling}

RHIC data offers some guidance for modelling cosmic-ray air showers.  Nuclear effects 
are significant; $AA$ collisions are not merely superpositions of $pp$ or even 
$pA$ collisions.  A blast-wave model does a good job of explaining soft particle
production; it may be of interest for simple simulations. Three
aspects of the RHIC data that may be particularly significant are:

1) The reduced forward particle production in $dA$ collisions 
may be due to decreased parton densities in nuclei at low $x$.  This may reduce 
number of high energy muons, and/or affect
the downward energy flow in
air showers.  More quantitative saturation models are needed to model higher
energy interactions.

2) The strangenss content in nuclear collisions is significantly increased over
$pp$ collisions even at moderate $N_{part}$, in $pA$ and $AA$, and, probably, in $\pi A$ collisions. 
At RHIC
energies, the increase is about 50\% for nitogren-nitrogen collisions.  
Since $K^\pm$ decay faster than $\pi^\pm$, in air showers, they are more likely to decay before 
interacting, so the increased strangeness should lead to more high energy muons.  The
increased strangeness will also reduce the $\pi^0$ fraction in collisions, slowing the conversion 
of hadronic energy into electromagnetic energy.  This may partially counterbalance the previous item.

3) Lighter systems may be modelled by using the $N_{part}$ dependence of heavy-ion systems.

\section{Conclusions and Future Prospects}

A  simple  explosive-exansion model can explain much
of the soft particle production data.  Similarly, a surface emission picture can 
explain a lot (but not all) of the high $p_T$ particle production.  However,
pQCD calculations do not reproduce the parameters required for these models.
It may be that a new non-perturbative QGP is being produced at RHIC energies.

Over the next few years, the RHIC detectors will upgrade their subsystems.
The major experimental goals are high-statustics studies of open charm 
and $\Upsilon$ production and of leptonic decays of vector mesons; the latter
may be sensitive to chiral symmetry breaking.  Another interesting idea is to
search for the tricritical point of QCD; this might be found at non-zero 
baryon density; RHIC can search for it point by scanning the machine energy. 

I thank the organizers for an enjoyable meeting, and my RHIC colleagues
for useful suggestions. This work was supported by the U.S. DOE under contract 
number DE-AC03076SF00098.

\end{document}